\documentclass[twocolumn,preprintnumbers,amsmath,amssymb,superscriptaddress]{revtex4}
\usepackage{graphicx}
\usepackage{dcolumn}
\usepackage{bm}
\usepackage{soul}
\usepackage{color}
\usepackage{epstopdf}
\usepackage[version=3]{mhchem}
\usepackage{lipsum}
\usepackage[outercaption]{sidecap}
\usepackage{floatrow}
\usepackage{hyperref}
\usepackage{appendix}

\begin{document}

\title{Efficient Analytical Approach for High-Pressure Melting Properties of Iron}
\author{Tran Dinh Cuong}
\email{cuong.trandinh@phenikaa-uni.edu.vn}
\affiliation{Phenikaa Institute for Advanced Study (PIAS), Phenikaa University, Hanoi 12116, Vietnam}
\author{Anh D. Phan}
\email{anh.phanduc@phenikaa-uni.edu.vn}
\affiliation{Phenikaa Institute for Advanced Study (PIAS), Phenikaa University, Hanoi 12116, Vietnam}
\affiliation{Faculty of Computer Science, Materials Science and Engineering, Artificial Intelligence Laboratory, Phenikaa University, Hanoi 12116, Viet Nam}
\date{\today}

\date{\today}

\begin{abstract}
Iron represents the principal constituent of the Earth's core, but its high-pressure melting diagram remains ambiguous. Here we present a simple analytical approach to predict the melting properties of iron under deep-Earth conditions. In our model, anharmonic free energies of the solid phase are directly determined by the moment expansion technique in quantum statistical mechanics. This basis associated with the Lindemann criterion for a vibrational instability can deduce the melting temperature. Moreover, we correlate the thermal expansion process with the shear response to explain a discontinuity of atomic volume, enthalpy, and entropy upon melting. Our numerical calculations are quantitatively consistent with recent experiments and simulations. The obtained results would improve understanding of the Earth's structure, dynamics, and evolution.
\end{abstract}

\maketitle


\section{INTRODUCTION}
The melting transition of iron has attracted considerable attention due to numerous applications in modern science and technology. For instance, since iron is the principal component of the Earth's core \cite{1}, its melting properties are used to consider the heat budget, the magnetic field generation, and the thermal evolution of our planet \cite{2,3,4}. Additionally, precise knowledge of molten iron can enhance the efficiency of advanced manufacturing processes, particularly as metal 3D printing \cite{5,6,7}.

In the past thirty years, several experimental approaches have been developed to measure the melting behaviors of iron under extreme conditions. One can directly determine the melting temperature of iron up to 200 GPa by laser-heated diamond anvil cell (LH DAC) experiments \cite{8}. Although LH DAC is a unique and powerful static compression technique, its reliability depends on many factors, including the melting criterion, the heating method, and the temperature metrology \cite{9,10}. Performing dynamic shock-wave (SW) experiments \cite{11,12,13,14,15} allows us to recreate an ultra-high pressure of 330 GPa at the Earth's inner core boundary (ICB). However, SW experiments \cite{11,12,13,14,15} can only indirectly provide a few limited melting points. Moreover, in the case of iron, shock data is appreciably higher than static data \cite{16}. At the ICB pressure, the deviation between SW and LH DAC extrapolations can be up to 3000 K \cite{17}. 

On the computational side, \textit{ab initio} molecular dynamics (AIMD) simulation and classical molecular dynamics (CMD) simulation are two primary techniques to estimate the melting properties of iron. AIMD and CMD can yield insights into melting mechanisms and expand the pressure range up to 1500 GPa \cite{18,19,20}. Nevertheless, employing AIMD and CMD requires heavy computational workloads \cite{21,22,23}. The simulation complexity grows dramatically with the number of atoms and phases. Furthermore, computational melting curves of iron show wide variations and are not sufficient to resolve experimental discrepancies \cite{8,16,17}. Besides, it is difficult to obtain a comprehensive description of the solid-liquid transition of iron-based alloys \cite{24}. Consequently, developing simulation methods and theoretical models for the melting phenomenon is still an intriguing problem in physics.

From a theoretical perspective, one can determine the melting boundary of iron and its alloys by the statistical moment method (SMM) \cite{25,26,27,28}. The SMM constructs a moment recurrence formula to calculate the anharmonic free energy of the solid phase \cite{29,30,31}. On that basis, the melting temperature is inferred from the spinodal condition \cite{32} and the dislocation-mediated melting theory \cite{33}. Although the SMM can give us the melting curve of iron within a few minutes, the melting gradient is overestimated at the high-pressure regime \cite{25,26,27,28}. Hoc \textit{et al.} \cite{34} have taken into account vacancy effects to obtain more accurate results for iron-carbon solid solutions. Unfortunately, this vacancy model is invalidated at 100 GPa due to ignorance of the vacancy formation volume \cite{35}. In recent works, Hieu \textit{et al.} \cite{36,37} have combined the SMM analysis \cite{38} with the Lindemann criterion \cite{39} to achieve a flatter melting curve for iron up to 150 GPa. Numerical calculations of Hieu \textit{et al.} \cite{36,37} are in good agreement with prior static experiments. However, the mechanical approach of Lindemann \cite{39} fails to explain an abrupt change in the atomic volume, enthalpy, and entropy upon melting \cite{33}.

Remarkably, Chuvildeev and Semenycheva \cite{40,41} have proposed a phenomenological model to obtain the fusion volume from the macroscopic properties of solids. The pressure difference at the solid-liquid interface is supposed to be analogous to the static Casimir effect \cite{42} in quantum field theory. From the ``Casimir force analog" (CFA), Chuvildeev and Semenycheva \cite{40,41} have found a simple correlation between the thermal expansion process and the transverse wave propagation upon melting. This correlation works well for a variety of transition metals. In addition, the atomic vibration amplitude extracted from the CFA model \cite{40,41} is quantitatively consistent with the Lindemann criterion \cite{39}. Nevertheless, this model has not yet been applied to the high-pressure regime. 

Here we combine the SMM model \cite{38} with the Lindemann criterion \cite{39} and the CFA approximation \cite{40,41} to predict the melting properties of iron up to 350 GPa. It is well known that iron has only three lattice forms, including body-centered cubic (BCC), face-centered cubic (FCC), and hexagonal close-packed (HCP) structures \cite{43}. Notwithstanding, its accurate phase diagram near the ICB is still under debate. Through AIMD simulations, Belonoshko \textit{et al.} \cite{44} have suggested that self-diffusion processes may stabilize the BCC phase at high temperatures. Liquid-like behaviors of BCC-iron \cite{45} are useful for explaining the seismic anisotropy \cite{46} and the low rigidity \cite{47} of the inner core. Unfortunately, the stability limit of BCC-iron remains enigmatic due to numerous challenges in free energy calculations \cite{48}. Size effects of supercells on entropies \cite{48} are not fully understood. On the other hand, the latest experiments \cite{49,50,51} have detected melting signatures of the HCP phase up to $\sim300$ GPa. This evidence makes HCP-iron a leading candidate for the Earth's core \cite{49,50,51}. Hence, for simplicity, we only carry out SMM analyses for the HCP structure. Our numerical results are comprehensively compared with previous studies.

\section{THEORETICAL BACKGROUND}
In the SMM, a HCP crystal is described by an assembly of identical anharmonic oscillators having the Hook constant $k$, the atomic mass $m$, the Einstein frequency $\omega_E$, and the nonlinear parameters $\beta$, $\gamma_1$, $\gamma_2$, and $\gamma$ \cite{52}. These quantities are determined by applying the Taylor series to the cohesive energy $u_0$ as
\begin{eqnarray}
k&=&\frac{1}{2}\sum_i\left(\frac{\partial^2u_0}{\partial u_{ix}^2}\right)_{eq},\quad\omega_E=\left(\frac{k}{m}\right)^{\frac{1}{2}},\nonumber\\
\beta&=&\frac{1}{2}\sum_i\left(\frac{\partial^3u_0}{\partial u_{ix}\partial u_{iy}^2}\right)_{eq},\quad\gamma_1=\frac{1}{48}\sum_i\left(\frac{\partial^4u_0}{\partial u_{ix}^4}\right)_{eq},\nonumber\\
\gamma_2&=&\frac{1}{8}\sum_i\left(\frac{\partial^4u_0}{\partial u_{ix}^2\partial u_{iy}^2}\right)_{eq},\quad\gamma=4\left(\gamma_1+\gamma_2\right),
\label{eq:1}
\end{eqnarray}
where $u_{ix}$ and $u_{iy}$ are displacements of the \textit{i}-th atom along $x$ and $y$ axes, respectively (see the Appendix A). Based on the moment expansion analysis in quantum statistical physics, the total vibrational free energy $\psi(V,T)$ of an atom can be straightforwardly computed by \cite{29,30,31}
\begin{eqnarray}
&\psi&(V,T)=\cfrac{1}{2}u_{0}+3\theta [X+\ln(1-e^{-2X})] \nonumber\\
&+&\frac{3\theta^2}{K^2}\left[\gamma_2X^2\coth^2X-\frac{2}{3}\gamma_1a_1 \right]\nonumber\\
&+&\frac{6\theta^3}{K^4}\left[\frac{4}{3}\gamma_2^2a_1X\coth{X}-2\gamma_1\left(\gamma_1+2\gamma_2\right)a_1\left(2a_1-1\right)\right]\nonumber\\
&+&\frac{3\theta^2\beta}{K}\left(\frac{2\gamma a_1}{3K^3}\right)^{\frac{1}{2}}+3\theta^3\beta\left[\left(\frac{2\gamma a_1}{3K^3}\right)^{\frac{3}{2}}+\frac{2k\gamma\beta}{K^6}\right],\nonumber\\
&a_1&=1+\frac{1}{2}X\coth{X},\quad K=k-\frac{\beta^2}{3\gamma},
\end{eqnarray}
where $V$ is the atomic volume, $T$ is the lattice temperature, $\theta=k_BT$ is the thermal energy, $k_B$ is the Boltzmann constant, $X=\hbar\omega_E/2\theta$ denotes quantum effects on the studied system, and $\hbar$ is the reduced Planck constant.

The hydrostatic pressure $P$ is defined by
\begin{eqnarray}
P=-\left(\frac{\partial\psi(V,T)}{\partial V}\right)_T=-\frac{a}{3V}\left(\frac{\partial\psi(a,T)}{\partial a}\right)_T,
\label{eq:3}
\end{eqnarray}
where $a=(V\sqrt{2})^{\frac{1}{3}}$ is the nearest neighbor distance. Solving Eq.(\ref{eq:3}) at $T=0$ K provides the value of $a(P,0)$. When $T>0$ K, one can consider the atomic arrangement by \cite{53}
\begin{eqnarray}
a(P,T)=a(P,0)+y(P,T),
\label{eq:4}
\end{eqnarray}
\begin{eqnarray}
y(P,T)&=&\sqrt{\frac{2\gamma\theta^2A}{3K^3}}-\frac{\beta}{3\gamma}+\nonumber\\
&+&\frac{1}{K}\left(1+\frac{6\gamma^2\theta^2}{K^4}\right)\left(\frac{1}{3}-\frac{2\beta^2}{27\gamma k}\right)\frac{\beta k}{\gamma},
\label{eq:5}
\end{eqnarray}
where $y(P,T)$ is the lattice expansion during heating. The explicit form of $A$ was reported in Ref.\cite{29} by employing a force balance condition for the tagged atom.

According to Lindemann's picture \cite{39}, the melting transition is driven by a vibrational instability of solids. Specifically, melting starts when the ratio of the root-mean-square vibration amplitude to the nearest neighbor distance reaches a critical value (the Lindemann parameter $\delta_L$) \cite{54}. This physical perspective allows us to derive an analytical expression for the melting temperature $T_m$ as \cite{55}
\begin{eqnarray}
\frac{T_m(P)}{T_m(P_0)}=\left[\frac{V(P)}{V(P_0)}\right]^{\frac{2}{3}}\left[\frac{\Theta_D(P)}{\Theta_D(P_0)}\right]^{2},
\label{eq:6}
\end{eqnarray}
where $\Theta_D\approx\cfrac{4\hbar}{3k_B}\omega_E$ is the Debye temperature \cite{56,57}. Conventionally, the reference melting point $T_m(P_0)$ is taken from experiment, and the right side of Eq.(\ref{eq:6}) is determined at room temperature \cite{58,59,60}.  

Within the CFA approximation \cite{40,41}, the fusion volume $\Delta V_m$ can be written by
\begin{eqnarray}
\frac{\Delta V_m(P)}{\Delta V_m(P_0)}&=&\left[\frac{\beta_T(P)}{\beta_T(P_0)}\right]^{\frac{4}{3}}\left[\frac{G(P)}{G(P_0)}\right]^{\frac{1}{2}}\left[\frac{T_m(P)}{T_m(P_0)}\right]^{\frac{1}{3}},
\label{eq:7}
\end{eqnarray}
where $\beta_T$ is the thermal expansion coefficient and $G$ is the shear modulus. Utilizing the SMM analysis gives \cite{61}
\begin{eqnarray}
\beta_T=\frac{1}{V}\left(\frac{\partial V}{\partial T}\right)_P,
\label{eq:8}
\end{eqnarray}
\begin{eqnarray}
G=\frac{K^5}{2\pi a(1+\nu)\left[K^4+2\gamma^2\theta^2a_1\left(2a_1-1\right)\right]},
\label{eq:9}
\end{eqnarray}
where $\nu$ is the Poisson ratio. Note that the Poisson ratio is almost unchanged with pressure and temperature \cite{33,62}. From these, one can estimate the fusion enthalpy $\Delta H_m$ and the fusion entropy $\Delta S_m$ by the Clausius-Clapeyron relation as \cite{63,64}
\begin{eqnarray}
\frac{\partial P}{\partial T_m}=\frac{1}{T_m}\frac{\Delta H_m}{\Delta V_m}=\frac{\Delta S_m}{\Delta V_m}.
\label{eq:10}
\end{eqnarray}

\section{RESULTS AND DISCUSSION}
To describe the interatomic interaction in transition metals, the SMM model typically adopts the Morse pair potential $\varphi_i$ as \cite{35}
\begin{eqnarray}
u_0=\sum_i\varphi_{i}=\sum_iD\left[e^{-2\alpha\left(r_{i}-r_0\right)}-2e^{-\alpha\left(r_{i}-r_0\right)}\right],
\label{eq:11}
\end{eqnarray}
where $D$ is the dissociation energy, $\alpha^{-1}$ is the decay length, $r_{i}$ is the distance between $i$-th and $0$-th atoms, and $r_0$ is the equilibrium value of $r_{i}$. The Morse parameters of iron are listed in Table \ref{tab:table1} by applying the extended X-ray absorption fine structure technique \cite{65}.

\begin{table}[htp]
\begin{ruledtabular}
\begin{tabular}{ccccc}
\textrm{$D$}&
\textrm{$\alpha$}&
\textrm{$r_0$}&
\textrm{$T_m(P_0)$}&
\textrm{$\Delta V_m(P_0)$}\\
(eV)&(\AA$^{-1}$)&(\AA)&(K)&(\AA$^3$)\\
\colrule
$0.42\pm0.12$\footnote{Taken from Ref.\cite{65}} &$1.40\pm0.20$$^a$ & 2.856$^a$ &3500\footnote{Taken from Ref.\cite{66}} & 0.19\footnote{Taken from Ref.\cite{20}}\\
\end{tabular}
\caption{\label{tab:table1} Thermodynamic parameters used to capture the high-pressure melting behaviors of iron \cite{20,65,66}. The FCC-HCP-liquid triple point at $P_0=100$ GPa \cite{66} is chosen as the reference melting point.}
\end{ruledtabular}
\end{table}

\begin{figure}[htp]
\includegraphics[width=9 cm]{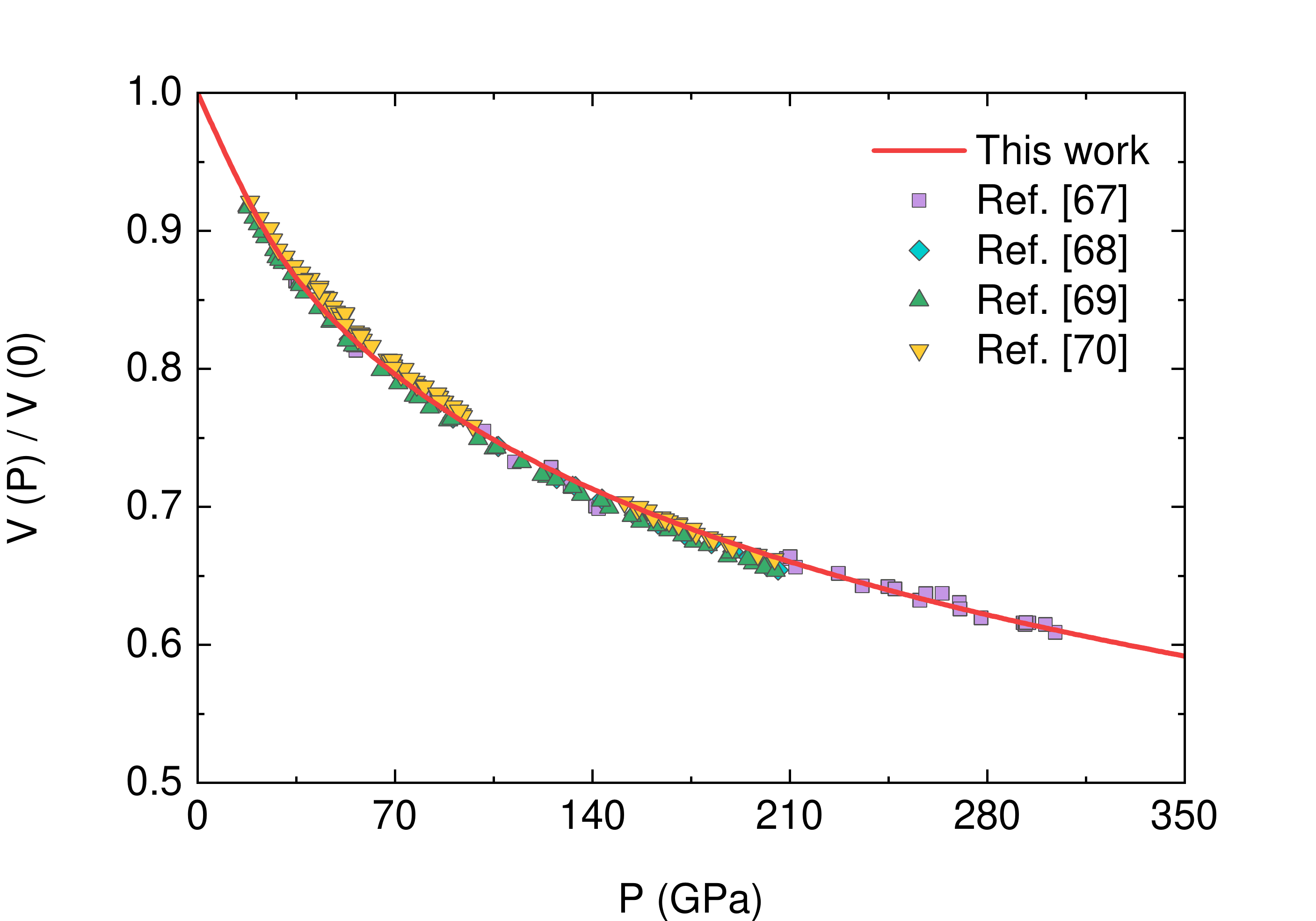}
\caption{\label{fig:1}(Color online) The equation of state of iron at room temperature obtained from the SMM analysis and DAC measurements \cite{67,68,69,70}.}
\end{figure}

Equation (\ref{eq:6}) reveals that the accuracy of the high-pressure melting curve depends crucially on the equation of state. Hence, we investigate isothermal compression effects on arrangement of iron atoms. As shown in Figure \ref{fig:1}, the normalized atomic volume $V(P)/V(0)$ reduces dramatically with pressure. At room temperature, when $P$ grows from 0 to 350 GPa, $V(P)/V(0)$ varies monotonically from 1 to 0.59. The maximum error between SMM calculations and DAC experiments \cite{67,68,69,70} is only $1.25$ \%. This value validates our analytical approach and the chosen interatomic potential. Our numerical data can be fitted by the Rydberg-Vinet equation of state as \cite{71}
\begin{eqnarray}
P&=&3K_0\left(\frac{V(P)}{V(0)}\right)^{-\frac{2}{3}}\left[1-\left(\frac{V(P)}{V(0)}\right)^{\frac{1}{3}}\right]\nonumber\\
&\times&\exp{\left\{\frac{3}{2}\left(K_0^{'}-1\right)\left[1-\left(\frac{V(P)}{V(0)}\right)^{\frac{1}{3}}\right]\right\}},
\label{eq:12}
\end{eqnarray}
where $K_0=159.06$ GPa and $K_0^{'}=5.87$ are the bulk modulus  and its pressure derivative under ambient conditions, respectively.

\begin{figure}[htp]
\includegraphics[width=9 cm]{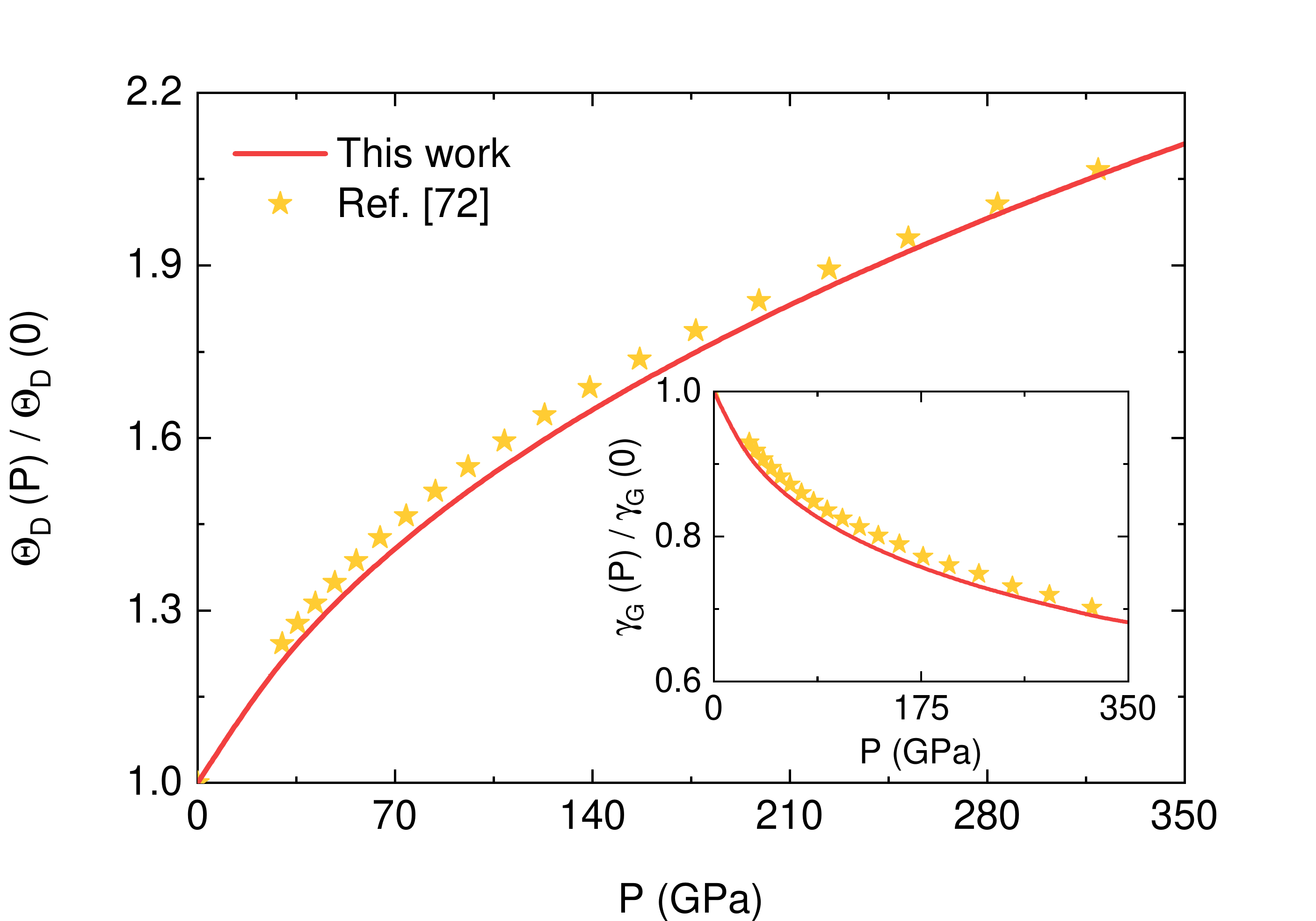}
\caption{\label{fig:2}(Color online) The scaled Debye temperature of iron as a function of pressure provided by the SMM and experiments \cite{72}. Inset: The pressure dependence of the normalized vibrational Grüneisen parameter.}
\end{figure}

Figure \ref{fig:2} shows a correlation between pressure and the scaled Debye temperature in Eq.(\ref{eq:6}). One can realize significant growth in $\Theta_D(P)/\Theta_D(0)$ with increasing $P$. Physically, the shrinkage of lattice space enhances the effective spring hardness of atomic oscillators \cite{57,58,59,60}. Hence, the scaled Debye temperature of iron can be up to 2.11 at 350 GPa. The compression sensitivity of $\Theta_D$ is reflected in the vibrational Grüneisen parameter $\gamma_G$, which is \cite{57,58,59,60}
\begin{eqnarray}
\gamma_G=-\frac{\partial\ln{\Theta_D}}{\partial\ln{V}}.
\label{eq:13}
\end{eqnarray}
A reduction in $\gamma_G$ can be quantitatively described by the well-known $q-$model as \cite{57,58,59,60}
\begin{eqnarray}
\frac{\gamma_G(P)}{\gamma_G(0)}=\left(\frac{V(P)}{V(0)}\right)^q,
\label{eq:14}
\end{eqnarray}
where $q=0.73$ is inferred from the SMM. Our predictions are in good accordance with previous experiments \cite{72} utilizing the Rietveld refinement \cite{73} of powder diffraction data. Overall, precise knowledge of $\Theta_D$ and $\gamma_G$ can suggest innovative ways of determining the melting temperature \cite{59,60}, the thermal equation of state \cite{69,70}, and the electrical resistivity \cite{74,75} of metals and alloys.

\begin{figure}[htp]
\includegraphics[width=9 cm]{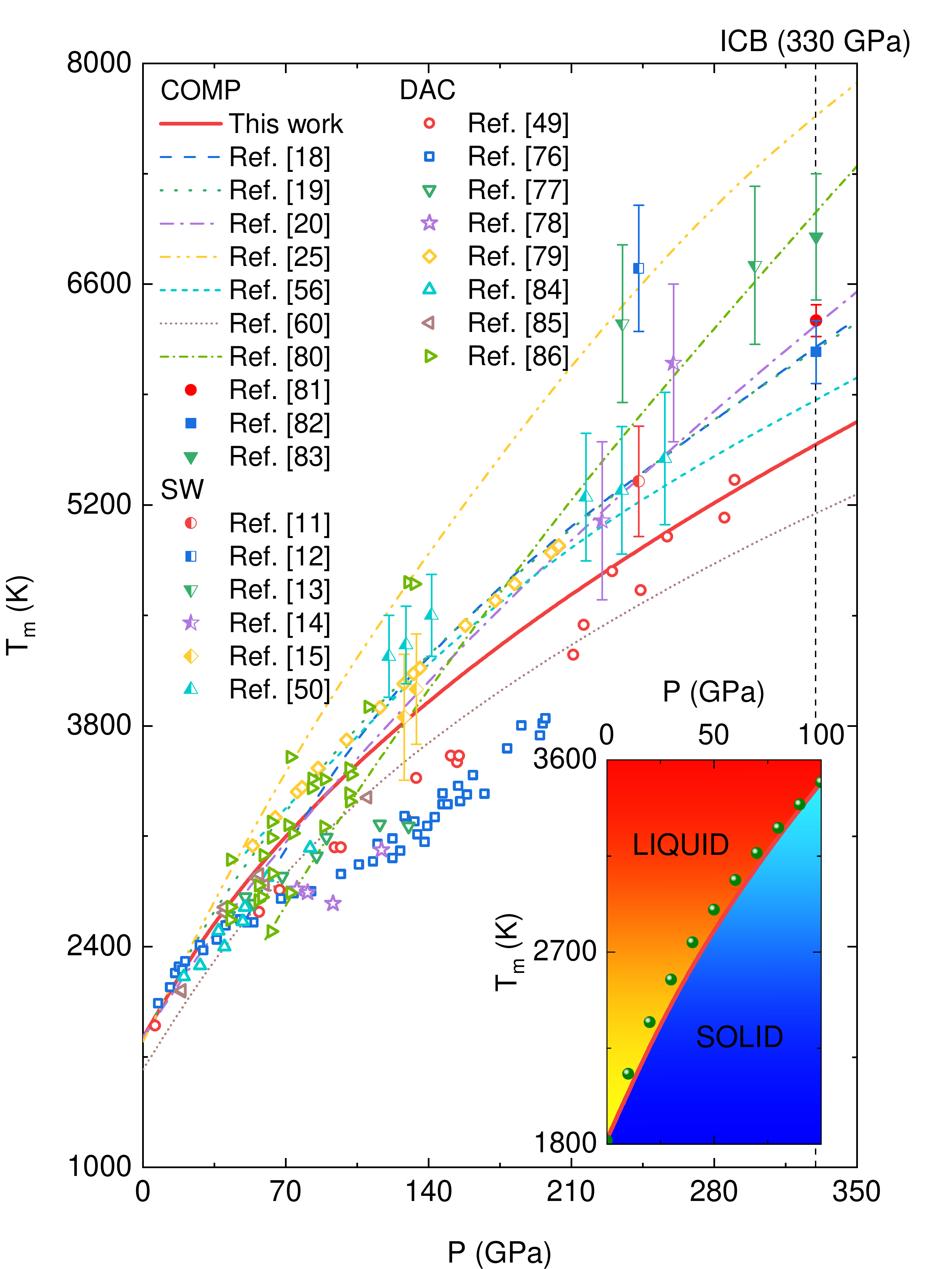}
\caption{\label{fig:3}(Color online) Pressure effects on the melting temperature of iron given by the SMM, experiments \cite{11,12,13,14,15,49,50,76,77,78,79,84,85,86}, and other computations \cite{18,19,20,25,56,60,80,81,82,83}. Inset: The solid-liquid boundary deduced from our analysis (red solid line) and an average between XRD, SMS, and XAS data \cite{86} (olive dotted line) up to 100 GPa.}
\end{figure}

Figure \ref{fig:3} shows how the melting temperature of iron depends on pressure. It is conspicuous that the existing data are alarmingly self-conflicting. Pioneering LH DAC works \cite{76,77} have suggested a shallow melting curve for iron. This melting tendency has been confirmed by recent static experiments adopting X-ray absorption spectroscopy (XAS) \cite{78}. In contrast, based on the fast X-ray diffraction (XRD) technique, Anzellini \textit{et al.} \cite{79} have found a dramatic growth in the melting temperature of iron. The difference between XRD and XAS data \cite{78,79} in LH DAC can be up to 1000 K at 135 GPa, which cannot be explained by experimental errors. The extrapolated melting line of Anzellini \textit{et al.} \cite{79} is consistent with molecular dynamics \cite{18,19,20,80,81,82} and quantum Monte Carlo \cite{83} simulations. However, this result \cite{79} is not well supported by synchrotron Mössbauer spectroscopy (SMS) studies \cite{84,85}. SMS data \cite{84,85} are scattered between XRD and XAS melting boundaries \cite{78,79}. These persisting discrepancies have raised a heated debate for decades.  

Remarkably, in recent work, Morard \textit{et al.} \cite{86} have reconciled the contradiction in LH DAC data \cite{76,77,78,79,84,85}. They have demonstrated that the anomalously low melting point in Ref.\cite{76,77,78} is due to chemical contaminations. Carbon atoms can migrate from the anvil surface into an iron sample to form carbides, e.g., cementite \cite{87}. This phenomenon decreases significantly the thermodynamic stability of iron \cite{88}. Besides, the inconsistency between XRD and SMS measurements \cite{79,84,85} arises from the thermal pressure overestimation. After recalculating the sample pressure, Morard \textit{et al.} \cite{86} obtained an excellent agreement between XRD and SMS results \cite{79,84,85}. These conclusions have been validated by the latest XAS experiments \cite{86}. Consequently, the phase boundary reported in Ref.\cite{86} may be very close to the real melting curve of iron.

As shown in the inset, we can accurately reproduce experimental data of Morard \textit{et al.} \cite{86} via the Lindemann criterion \cite{39}. The deviation between these approaches is smaller than 3.95 $\%$ in a wide pressure region $0\leq P\leq 100$ GPa. In the quasi-harmonic approximation \cite{38}, the Lindemann parameter of iron is $\delta_L=\sqrt{3k_BT_m/ka^2}=0.091$, which is typical for HCP metals (0.08-0.11) \cite{89}. Note that Eq.(\ref{eq:6}) only uses some basic thermodynamic properties at room temperature to investigate the solid-liquid transition. Therefore, the superheating problem in previous SMM studies \cite{25,26,27,28} can be adequately resolved without heavy computational workloads. 

Surprisingly, for $P>100$ GPa, our theoretical estimates are lower than SW data of Li \textit{et al.} \cite{50} by only 10 \%. According to Li \textit{et al.} \cite{50}, the prior SW information \cite{11,12,13,14,15} is very uncertain due to the limitation of optical pyrometers, the porosity of thin iron samples, and the imperfection of ``sample/window" interfaces. Through time‐resolved pyrometers, Li \textit{et al.} \cite{50} have essentially achieved a consensus between dynamic and static compression processes. This notable finding \cite{50} has imposed relatively tight constraints on the melting behaviors of iron in the deep Earth's interior.

To facilitate planetary modeling, we fit present SMM results by the Simon-Glatzel equation as \cite{90}
\begin{eqnarray}
T_m=T_{m}^{*}\left(\frac{P}{P^*}+1\right)^{c},
\label{eq:15}
\end{eqnarray}
where $T_{m}^{*}=1822.88$ K, $P^*=32.51$ GPa, and $c=0.4644$ are comparable to other semi-empirical calculations \cite{56,60}. From Eq.(\ref{eq:15}), we obtain $T_m=5586$ K at 330 GPa. This value agrees quantitatively well with $T_m=5500\pm 220$ K predicted by the resistance-heated diamond anvil cell (RH DAC) method \cite{49}. In contrast to LH DAC \cite{76,77,78,79,84,85,86} and SW \cite{11,12,13,14,15,50}, modern RH DAC measurements \cite{49} can directly yield the melting point of iron near the ICB pressure. Additionally, a relatively low thermal gradient in RH DAC samples gives confidence for the melting detection \cite{49}. Thus, $T_m=5586$ K may be a reasonable upper limit for the ICB temperature. On that basis, one can effectively determine geothermal profiles and internal dynamics of the Earth's core via an adiabatic model \cite{91} (see the Appendix B). 

\begin{figure}[htp]
\includegraphics[width=9 cm]{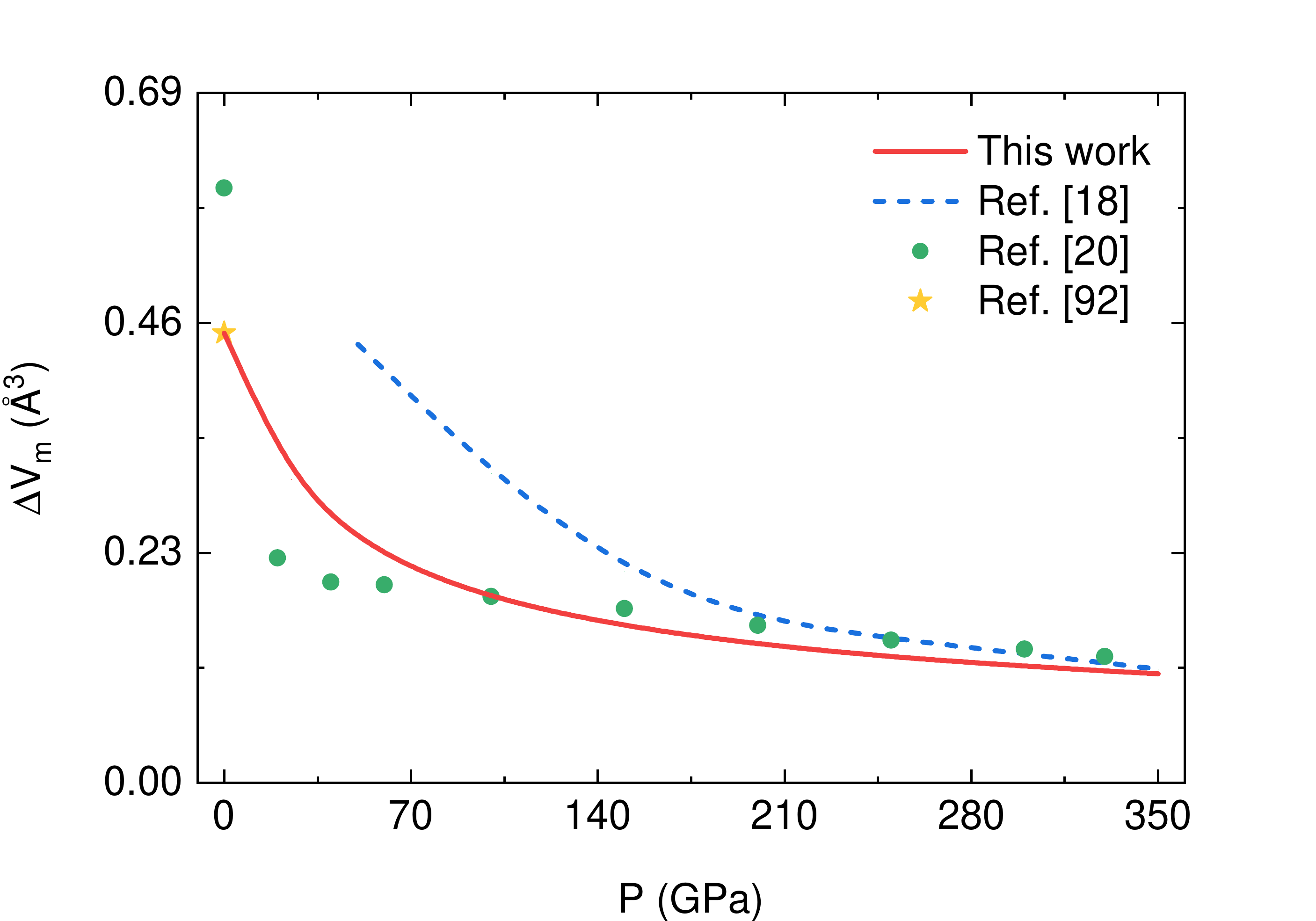}
\caption{\label{fig:4}(Color online) The fusion volume of iron versus pressure predicted by the SMM, AIMD \cite{18}, CMD \cite{20}, and experiments \cite{92}.}
\end{figure}

Figure \ref{fig:4} shows the fusion volume of iron as a function of pressure. According to the SMM analysis, $\Delta V_m$ drops quickly by a factor of 3.23 between 0 and 200 GPa. Then, $\Delta V_m$ becomes insensitive to compressions and saturates to $0.11$ \AA$^3$ at 350 GPa. The accuracy of the CFA approximation \cite{40,41} is comparable to the most sophisticated computational techniques \cite{18,20}, particularly under deep-Earth conditions. The small discrepancies between our one-phase approach and other simulations \cite{18,20} can be resolved by taking into account electronic excitations \cite{82} and the pressure dependence of interatomic potential parameters \cite{93}. 

\begin{figure}[htp]
\includegraphics[width=9 cm]{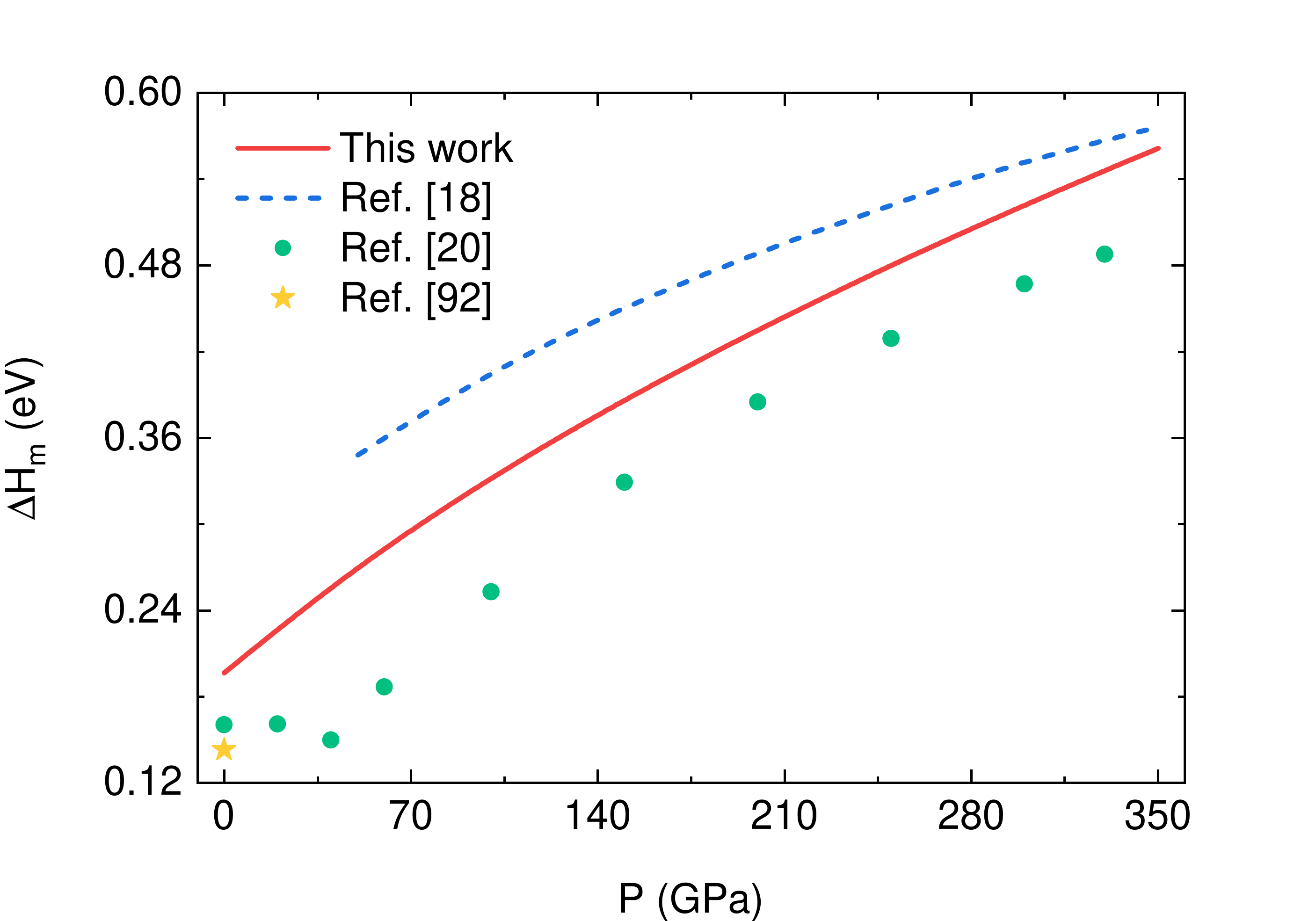}
\caption{\label{fig:5}(Color online) Correlations between pressure and the fusion enthalpy of iron given by the SMM, AIMD \cite{18}, CMD \cite{20}, and experiments \cite{92}.}
\end{figure}

Figure \ref{fig:5} shows compression effects on the fusion enthalpy of iron. It is clear to see that $\Delta H_m$ and $T_m$ vary with the same trend. Hence, from Eq.(\ref{eq:10}), the fusion entropy only slightly fluctuates along the solid-liquid boundary. At the ICB pressure, SMM calculations provide $\Delta H_m=0.54$ eV ($\Delta S_m=1.12k_B$), which is quite close to $\Delta H_m=0.57$ eV ($\Delta S_m=1.05 k_B$) and $\Delta H_m=0.48$ eV ($\Delta S_m=0.87k_B$) extracted from AIMD and CMD simulations \cite{18,20}, respectively. In practice, these findings can be applied to consider the Earth's total heat budget \cite{2} or optimize laser materials processing \cite{94}.

\section{CONCLUSION}

We have extended the SMM model to obtain a more comprehensive picture of the melting phenomenon of iron up to 350 GPa. Based on the Lindemann criterion, we have successfully reproduced the high-pressure melting diagram of Morard \textit{et al.} in Ref.\cite{86}. The upper bound for the ICB temperature has been estimated to be $T_m=5586$ K, which is in excellent accordance with $T_m=5500\pm220$ K taken from the latest RH DAC experiments. Furthermore, an abrupt change in atomic volume, enthalpy, and entropy upon melting has been fully described by the CFA approximation. Therefore, our numerical results would be useful for geodynamic modeling of planetary cores. It is possible to develop our one-phase approach to investigate the solid-liquid transition in low-dimensional and multicomponent systems.

\begin{acknowledgements}
T. D. Cuong is deeply grateful for the research assistantship at Phenikaa University.
\end{acknowledgements}

\appendix
\section{Crystalline Parameters}
If $i$-th and $0$-th atoms primarily interact with each other by the pair potential $\varphi_i(r_i)$, the analytical expression for crystalline parameters $k$, $\beta$, and $\gamma$ \cite{52} can be rewritten by
\begin{eqnarray}
k&=&\frac{1}{2}\sum_i\left(\frac{\partial^2\varphi_i}{\partial u_{ix}^2}\right)_{eq},\quad\beta=\frac{1}{2}\sum_i\left(\frac{\partial^3\varphi_i}{\partial u_{ix}\partial u_{iy}^2}\right)_{eq},\nonumber\\
\gamma&=&\frac{1}{12}\sum_i\left[\left(\frac{\partial^4\varphi_i}{\partial u_{ix}^4}\right)_{eq}+6\left(\frac{\partial^4\varphi_i}{\partial u_{ix}^2\partial u_{ix}^4}\right)_{eq}\right].
\label{eq:A1}
\end{eqnarray}
Utilizing the lattice theory of Leibfried and Ludwig \cite{95}, we have 
\begin{eqnarray}
\left(\frac{\partial^2\varphi_i}{\partial u_{ix}^2}\right)_{eq}=\left(\hat{o}^2\varphi_i\right)x_i^2+\left(\hat{o}\varphi_i\right),\nonumber
\end{eqnarray}
\begin{eqnarray}
\left(\frac{\partial^3\varphi_i}{\partial u_{ix}\partial u_{iy}^2}\right)_{eq}=\left(\hat{o}^3\varphi_i\right)x_iy_i^2+\left(\hat{o}^2\varphi_i\right)x_i,\nonumber
\end{eqnarray}
\begin{eqnarray}
\left(\frac{\partial^4\varphi_i}{\partial u_{ix}^4}\right)_{eq}=\left(\hat{o}^4\varphi_i\right)x_i^4+6\left(\hat{o}^3\varphi_i\right)x_i^2+3\left(\hat{o}^2\varphi_i\right),\nonumber
\end{eqnarray}
\begin{eqnarray}
\left(\frac{\partial^4\varphi_i}{\partial u_{ix}^2\partial u_{iy}^2}\right)_{eq}&=&\left(\hat{o}^4\varphi_i\right)x_i^2y_i^2+\left(\hat{o}^3\varphi_i\right)x_i^2+\nonumber\\
&+&\left(\hat{o}^3\varphi_i\right)y_i^2+\left(\hat{o}^2\varphi_i\right),
\label{eq:A2}
\end{eqnarray}
where $x_i$ and $y_i$ are Cartesian coordinates of the $i$-th atom. The operator $\hat{o}$ is defined by
\begin{eqnarray}
\hat{o}=\frac{1}{r_i}\frac{\partial}{\partial r_i}.
\label{eq:A3}
\end{eqnarray}
For a given structure, $x_i$, $y_i$, and $r_i$ are represented through the nearest neighbor distance $a$. Consequently, after knowing $a(P,T)$ (see Eqs.(\ref{eq:3})-(\ref{eq:5})), we can deduce $k$, $\beta$, and $\gamma$ at various thermodynamics conditions.

\section{The Earth's Temperature Profile}
\begin{figure}[htp]
\includegraphics[width=9 cm]{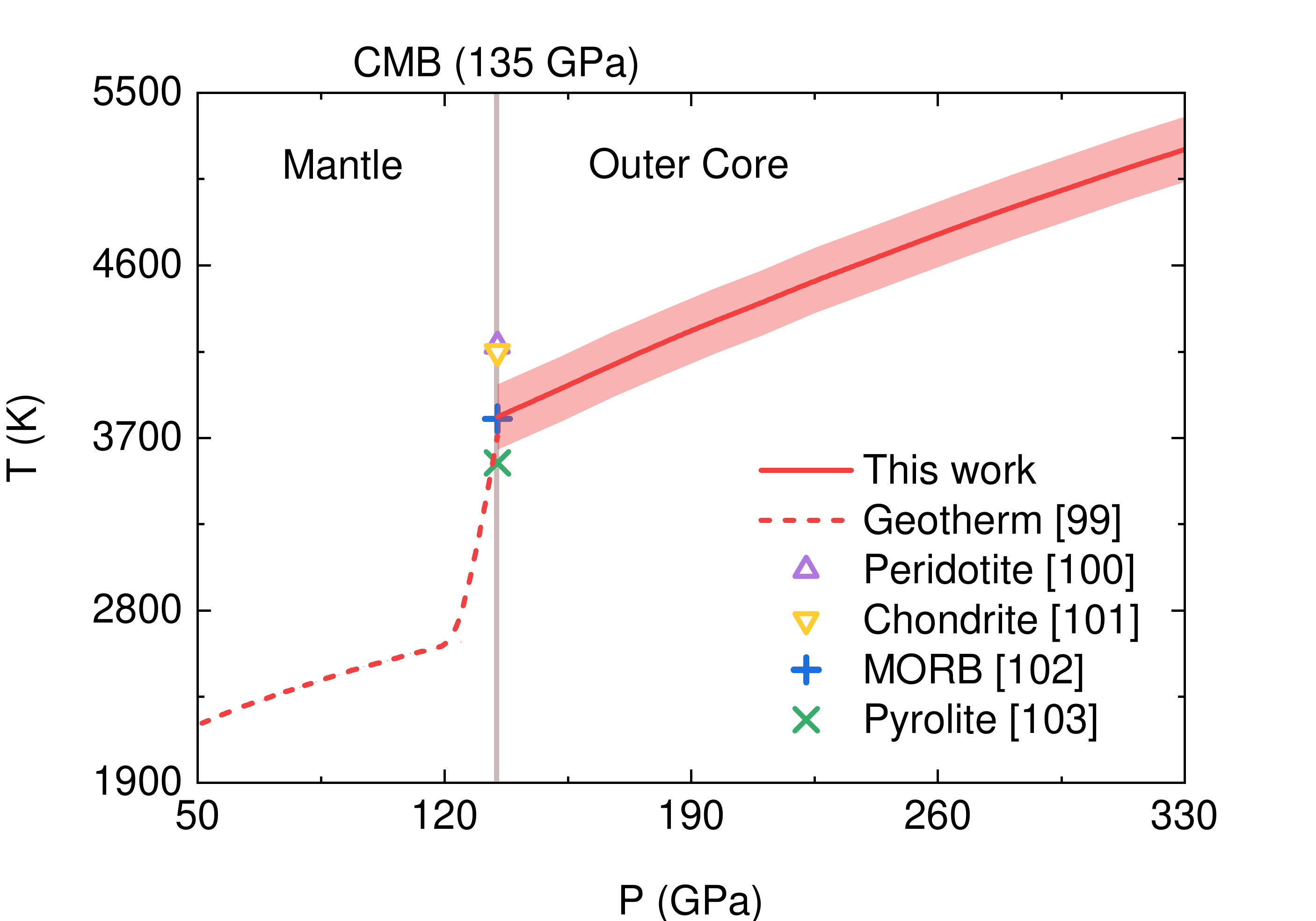}
\caption{\label{fig:6}(Color online) The Earth's temperature profile inferred from Eq.(\ref{eq:B2}) and Ref.\cite{99}. The solidus temperature of candidate lower-mantle materials \cite{100,101,102,103} is plotted for comparison.}
\end{figure}

According to seismological observations, the Earth's core is $\sim10$ $\%$ less dense than pure iron \cite{1}. This density deficit implies that our planet must include lighter elements, such as Si, O, S, C, and H \cite{1}. Physically, the presence of light impurities can depress the melting point of iron by at least $\Delta T=380\pm170$ K in the high-pressure region \cite{49}. Thus, we can approximate the ICB temperature $T_{ICB}$ by
\begin{eqnarray}
T_{ICB}=T_m-\Delta T=5206\pm170\ \textrm{K}.
\label{eq:B1}
\end{eqnarray}
From Eq.(\ref{eq:B1}), an adiabatic thermal profile inside the liquid outer core is given by \cite{96}
\begin{eqnarray}
T=T_{ICB}\left(\frac{\rho}{\rho_{ICB}}\right)^{\gamma_{th}},
\label{eq:B2}
\end{eqnarray}
where $\gamma_{th}=1.51$ is the thermodynamic Grüneisen parameter \cite{97}, $\rho$ is the core density derived from the Preliminary Reference Earth Model \cite{98}, and $\rho_{ICB}$ is the critical value of $\rho$ at the ICB. Numerical predictions based on Eq.(\ref{eq:B2}) are shown in Figure \ref{fig:6}.  

At the core-mantle boundary (CMB), using Eq.(\ref{eq:B2}) yields $T_{CMB}=3809\pm170$ K, which is appreciably lower than the solidus temperature $T_s\approx4150\pm150$ K of peridotitic and chondritic assemblages \cite{100,101}. This result reveals that the lowermost mantle has escaped global melting, at least since the early Proterozoic Eon \cite{49}. Notably, our calculations suggest that the CMB temperature can be higher than $T_s=3800\pm150$ K of mid-oceanic ridge basalt \cite{102} and $T_s=3570\pm200$ K of pyrolite with 400 ppm water \cite{103}. Hence, these materials may be partially molten and cause seismic ultra-low velocity zones above the CMB \cite{104}. Overall, our new constraints on $T_{CMB}$ would improve understanding of phase relations and element partitioning at the base of the mantle.

\end{document}